\documentclass[prl,twocolumn,showpacs]{revtex4}
\usepackage{graphicx}

\begin{document}

\title{Statistics of Dynamics of Localized Waves}

\author{A.A.~Chabanov and A.Z.~Genack}

\affiliation{Physics Department, Queens College of CUNY, Flushing, NY 11367}

\date{June, 2001}

\begin{abstract}
The measured distribution of the single-channel delay time of localized microwave radiation and its correlation with intensity differ sharply from the behavior of diffusive waves. The delay time is found to increase with intensity, while its variance is inversely proportional to the fourth root of the intensity. The distribution of the delay time weighted by the intensity is found to be a double-sided stretched exponential to the 1/3 power centered at zero. The correlation between dwell time and intensity provides a dynamical test of photon localization.
\end{abstract}


\maketitle

Large fluctuations are a distinctive feature of transport in quantum and classical mesoscopic systems. Though the focus of mesoscopic physics in all its varieties has been almost exclusively on steady-state propagation \cite{Mesobook,ShengBook}, it is natural to view transport from a dynamical perspective. Indeed, the fundamental dimensionless ratio in the study of propagation and localization, the Thouless number $\delta$ \cite{Thouless}, is the ratio of two dynamical parameters, the level width $\delta\nu$ and the level spacing $\Delta\nu$, $\delta=\delta\nu/\Delta\nu$. The level width is the inverse of the Thouless time, which is the transit time through the sample, and the level spacing is the inverse of the Heisenberg time, which is the time required to explore all coherence volumes of the sample. The statics and dynamics of transport are closely related since, in the absence of inelastic processes, $\delta$ is equal to the dimensionless conductance $g$ \cite{Gang4}, which is the inverse of the degree of long-range intensity correlation \cite{Stephen,Feng,Walter}.

Whereas static aspects of transport are associated with the amplitude of the wave, dynamics is reflected in the phase \cite{Eisenbud,Wigner,Smith,Sebbah}. The single-channel delay time for a transmitted pulse in mode $b$ for incident mode $a$ is the first temporal moment of the pulse. In the limit of vanishing pulse bandwidth, the single-channel delay time is given by $\tau_{ab}(\omega)=d\phi_{ab}/d\omega\equiv\phi_{ab}^{\prime}$, where $\phi_{ab}$ is the phase and $\omega$ is the angular frequency \cite{Sebbah}. The pulse transmission coefficient in this limit is equal to the static transmission coefficient or intensity $I_{ab}(\omega)$. The configuration or space-averaged delay time is obtained by weighting the delay time with the intensity, $W_{ab}= I_{ab}\phi^{\prime}_{ab}$. When averaged over all input and output channels, this is the Wigner delay time \cite{Smith}, which is proportional to the density of states. In diffusive limit, the conditional probability distribution of the single-channel delay time normalized to its ensemble average, $\widehat{\phi^{\prime}}\equiv\phi^{\prime}_{ab}/ \langle\phi^{\prime}_{ab}\rangle$, for fixed normalized intensity, $\widehat{I}\equiv I_{ab}/\langle I_{ab}\rangle$, is a gaussian with variance inversely proportional to $\widehat{I}$ \cite{Azi,Bart}. Surprisingly, measurements of the distribution and correlation function of the single-channel delay time for diffusing microwave radiation \cite{Azi} were found to be in excellent agreement with the theory, even in samples with a considerable degree of long-range intensity correlation. But, the statistics of dynamics of localized waves must differ fundamentally from those for diffusing waves since the transmission spectrum appears as a series of narrow spikes, reflecting the condition, $\delta\nu<\Delta\nu$. Unlike diffusive waves, for which the delay time and intensity are uncorrelated \cite{Azi}, long delay times for localized waves are associated with peaks in the transmitted intensity, associated with resonant tunneling through localized states. Calculations of novel statistics of dynamics in localized media have been carried out for reflection of acoustic waves in one-dimensional systems, such as the earth's crust \cite{Sheng}, for reflection of electromagnetic radiation in quasi-one-dimensional media \cite{Beenakker}, and for electron transport in a one-dimensional potential \cite{Texier}. Statistical aspects of scattering have also been considered in chaotic cavities \cite{Muga,Seba,Brouwer,Fyodorov,Gopar} and in nuclear and atomic scattering \cite{Smith}.

In this Letter, we present the first measurements of the dynamic statistics of localized waves. The statistics of dynamics of microwave radiation within a window of localization in samples of randomly positioned dielectric spheres are compared to those for extended waves in a different frequency interval. We find that the distribution of the single-channel delay time becomes markedly asymmetric, while the distribution of the delay time weighted by intensity becomes extraordinarily broad. It falls as a stretched exponential to the power 1/3 instead of exponentially, as predicted for diffusing waves. The average normalized delay time at fixed intensity, $\langle \widehat{\phi^{\prime}}\rangle_{\widehat{I}}\,$, which is independent of $\widehat{I}$ for diffusive waves, is found to increase with $\widehat{I}$. At the same time, the decrease with intensity of the variance of the normalized delay time is substantially reduced. As a result, the delay time and intensity are correlated and afford a dynamical test for localization. The deviations from diffusive behavior are consistent with expectations for resonant transmission through localized modes.

We have measured the microwave field transmission coefficient $\sqrt{I_{ab}}\exp(i\phi_{ab})$ in ensembles of randomly positioned alumina spheres. The amplitude $\sqrt{I_{ab}}$ and the phase $\phi_{ab}$ of the field at the output surface, referenced to the field at the input surface, are obtained using a Hewlett-Packard HP8722C vector network analyzer. Alumina spheres of diameter 0.95 cm and dielectric constant 9.86 \cite{PRL} are contained in a 7.3-cm-diameter copper tube at a volume fraction of 0.068. This low density is achieved by embedding the alumina spheres in 1.9-cm-diameter Styrofoam spheres with dielectric constant 1.08. The measurements are carried out in samples of length 49, 65, and 90 cm.

The degree of localization in these samples is given by the variance of the total transmission normalized to its ensemble average value, $var(s_{a})$, where $s_{a}=\Sigma_{b}I_{ab}/\langle\Sigma_{b}I_{ab}\rangle$ \cite{Nature,PRL}. At a threshold value of order unity, $var(s_{a})$ crosses over from a monotonic increase for extended waves to an exponential for localized waves, as the sample length $L$ increases \cite{Nature}. Calculated values of $var(s_{a})$, as well as measurements of scaling of $var(s_{a})$ \cite{Nature} and $\langle I_{ab}\rangle$ \cite{Unpub} with sample length in identical alumina samples, have allowed us to establish that the wave is localized in a narrow frequency range centered at $f\approx 10$ GHz, slightly above the first Mie resonance of the alumina spheres. Outside this range, the wave is extended. Here we compare the dynamic statistics in the frequency interval of 9.94-10.1 GHz, within the window of localization, to those for extended waves in the interval of 16.9-17 GHz near the fourth Mie resonance of the alumina spheres. These frequency intervals are sufficiently narrow that within them propagation parameters are nearly constant. 

The probability distributions $P(\widehat{\phi^{\prime}})$ of the normalized single-channel delay time in a sample of $L = 90$ cm are shown in Fig.~1. The ensemble-averaged values of $\phi^{\prime}_{ab}$ within the lower and upper frequency intervals are 122 and 120 ns, respectively. The values of $var(s_a)$ at these frequencies are, respectively, 7.1 and 0.37, indicating that radiation is localized in the lower frequency interval and extended in the upper interval, though the intensity correlation is high. The distribution $P(\widehat{\phi^{\prime}})$ for extended waves is compared to that in the diffusive limit \cite{Bart},
\begin{equation}
P\left(\widehat{\phi^{\prime}}\right)={1\over 2}{Q \over\left[Q+(\widehat{\phi^{\prime}}-1)^2\right]^{3/2}}. 
\label{}
\end{equation}
The parameter $Q$ is a function only of the ratio of $L$ and the diffusive absorption length $L_{a}$ \cite{Bart}. Measurements of the scaling of $\langle I_{ab}\rangle$ with sample length for extended waves yield $L_a = 21.3\pm 1.6$ cm \cite{Unpub}, which translates into $Q=0.215$. This value is substituted in Eq.~(1) to produce the curve in Fig.~1. We note, however, that a fit of Eq.~(1) to the data using $Q$ as a fitting parameter gives $Q=0.249$, which corresponds to $L_{a}=26.0\pm 0.4$ cm. The underestimate of absorption, resulting from the fit of Eq.~(1), indicates the beginning of a breakdown of the diffusion theory of \cite{Bart}, which is expected in strongly correlated samples. In contrast, the normalized delay time distribution for localized waves bears little resemblance to the predictions of diffusion theory. It is asymmetrical and reaches its peak at a value of $\widehat{\phi^{\prime}}$ below its average value of unity.

\begin{figure}
\includegraphics[width=\columnwidth]{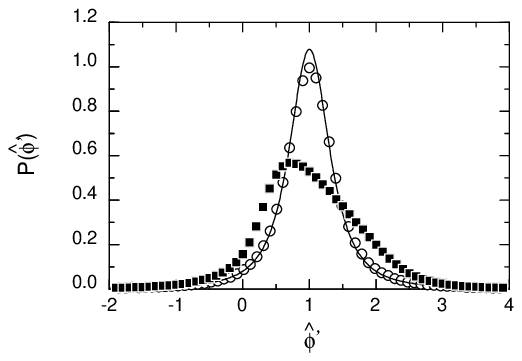}
\caption{Probability distribution of the normalized delay time for extended (circles) and localized (squares) waves in a sample of $L = 90$ cm. The curve is the distribution $P(\widehat{\phi^{\prime}})$ in the diffusive limit \cite{Bart}, given by Eq.~(1) with $Q = 0.215$.}
\end{figure}

We find further that the conditional probability distribution $P_{\widehat{I}}(\widehat{\phi^{\prime}})$ for a fixed value of $\widehat{I}$ for extended waves is well described by a gaussian for any $\widehat{I}$ and $L$, in agreement with the diffusion theory. The variation of $\langle\widehat{\phi^{\prime}}\rangle_{\widehat{I}}$ and $var(\widehat{\phi^{\prime}})_{\widehat{I}}$ upon $\widehat{I}$ in a sample of $L=90$ cm is shown in Fig.~2. Whereas $\langle\widehat{\phi^{\prime}}\rangle_{\widehat{I}}$ is, as expected, nearly independent of $\widehat{I}$ (Fig.~2a), $var(\widehat{\phi^{\prime}})_{\widehat{I}}$ shows a departure from the prediction for diffusive waves of $Q/2\widehat{I}$, for $\widehat{I}>0.5$ (Fig.~2b). This is consistent with the deviation of the distribution $P(\widehat{\phi^{\prime}})$ for extended waves from diffusion theory, seen in Fig.~1. 

\begin{figure}
\includegraphics[width=\columnwidth]{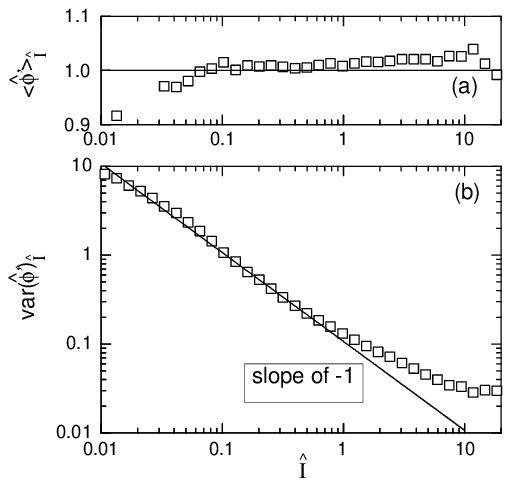}
\caption{Variation with $\widehat{I}$ of (a) $\langle\widehat{\phi^{\prime}}\rangle_{\widehat{I}}$ and (b) $var(\widehat{\phi^{\prime}})_{\widehat{I}}$ for extended waves for $L=90$ cm. The lines in (a) and (b) are the diffusive limits \cite{Bart}, $\langle\widehat{\phi^{\prime}}\rangle_{\widehat{I}}=1$ and $var(\widehat{\phi^{\prime}})_{\widehat{I}}=Q/2\widehat{I}$, respectively, with $Q=0.215$.}
\end{figure}

In contrast to the gaussian distribution found for extended waves, the conditional probability distribution $P_{\widehat{I}}(\widehat{\phi^{\prime}})$ for localized waves exhibits an exponential fall-off, with an asymmetry in the distribution, which increases with decreasing $\widehat{I}$ (Fig.~3). The variation of $\langle\widehat{\phi^{\prime}}\rangle_{\widehat{I}}$ and $var(\widehat{\phi^{\prime}})_{\widehat{I}}$ with $\widehat{I}$ for different values of $L$ is presented in Fig.~4. As seen in Fig.~4a, $\langle\widehat{\phi^{\prime}}\rangle_{\widehat{I}}$ markedly increases with $\widehat{I}$ to an extent, which increases with $L$. The variation of $var(\widehat{\phi^{\prime}})_{\widehat{I}}\,$, seen in Fig.~4b, is even more striking. For all sample lengths, we find that for $\widehat{I}>0.5$, $var(\widehat{\phi^{\prime}})_{\widehat{I}}$ converges to $q/(\widehat{I})^{1/4}$, with $q=0.4$, shown as the line in Fig.~4b. This universal behavior at large $\widehat{I}$ suggests a similarity in the dynamics of localized and prelocalized states \cite{Preloc}. For smaller values of $\widehat{I}$, $var(\widehat{\phi^{\prime}})_{\widehat{I}}$ becomes smaller and falls more slowly with $\widehat{I}$, as sample length increases. For $L=90$ cm, $var(\widehat{\phi^{\prime}})_{\widehat{I}}=0.4/(\widehat{I})^{1/4}$ for any $\widehat{I}$.

\begin{figure}
\includegraphics[width=\columnwidth]{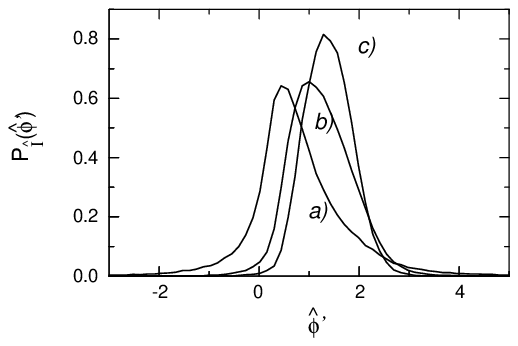}
\caption{Conditional probability distribution $P_{\widehat{I}}(\widehat{\phi^{\prime}})$ for the values of $\widehat{I}$ of 0.04 (a), 0.4 (b), and 4.0 (c) for localized waves for $L=90$ cm.}
\end{figure}

\begin{figure}
\includegraphics[width=\columnwidth]{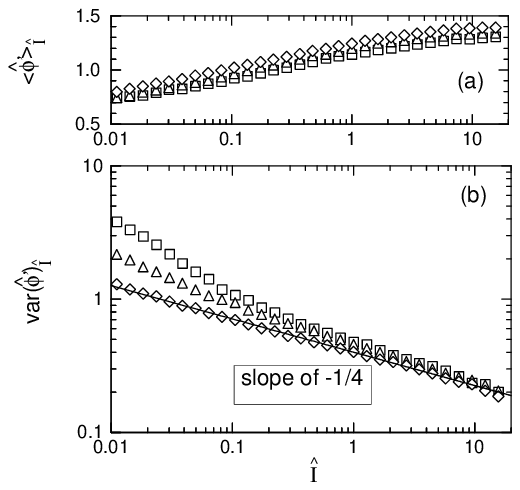}
\caption{Variation with $\widehat{I}$ of (a) $\langle\widehat{\phi^{\prime}}\rangle_{\widehat{I}}$ and (b) $var(\widehat{\phi^{\prime}})_{\widehat{I}}$ for localized waves for $L=49$ cm (squares), 65 cm (triangles), and 90 cm (diamonds). The line in (b) is $var(\widehat{\phi^{\prime}})_{\widehat{I}}=q/(\widehat{I})^{1/4}$, with $q=0.4$.}
\end{figure}

The probability distributions of the normalized weighted delay time, $\widehat{W}\equiv W_{ab}/\langle W_{ab}\rangle$, for extended and localized waves for $L=90$ cm are compared in Fig.~5. The dashed line is the predicted double-sided exponential distribution in the diffusive limit \cite{Bart}, with $Q = 0.215$, found in the upper frequency interval. The distribution for extended waves, however, is seen to deviate significantly from diffusion theory, indicating the sensitivity of this distribution to approaching localization. For localized waves, the distribution is considerably broader. It is well approximated by a double-sided stretched exponential, $P(\widehat{W})=a\exp(-b\vert\widehat{W}\vert^{1/3})$, with $a=0.44$ and $b=2.42$ for $\widehat{W}>0$, and $a=0.07$ and $b=5.50$ for $\widehat{W}<0$, shown as the solid line in Fig.~5. The distribution of $\widehat{W}$ is wider than that of $\widehat{I}$, reflecting the enhanced probability of long dwell times and the increased variance of dwell times at large values of the intensity for localized waves. 

\begin{figure}
\includegraphics[width=\columnwidth]{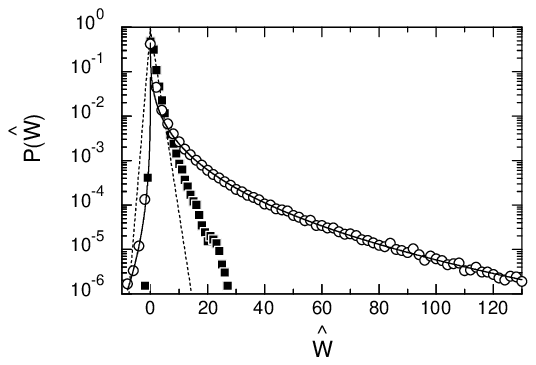}
\caption{Probability distribution of the normalized weighted delay time for extended (squares) and localized (circles) waves for $L=90$ cm. The dashed line is the distribution $P(\widehat{W})$ in the diffusive limit \cite{Bart}, with $Q=0.215$. The solid line is the model distribution, $P(\widehat{W})=a\exp(-b\vert\widehat{W}\vert^{1/3})$, with $a=0.44$ and $b=2.42$ for $\widehat{W}>0$, and $a=0.07$ and $b=5.50$ for $\widehat{W}<0$.}
\end{figure}

We have previously shown that $var(\widehat{I})$, as well as $var(s_{a})$, serve as indicators of localization, even in the presence of absorption \cite{PRL,Nature}. We find that the interaction between dynamic and static statistics associated with $\widehat{\phi^{\prime}}$ and $\widehat{I}$, respectively, may also be used to identify the range of localization. The correlation between $\widehat{I}$ and $\widehat{\phi^{\prime}}$ can be expressed as the dimensionless ratio, $\langle\widehat{I}\,\widehat{\phi^{\prime}}\rangle\equiv\langle I_{ab}\phi^{\prime}_{ab}\rangle/\langle I_{ab}\rangle\langle\phi^{\prime}_{ab}\rangle$. The frequency variation of this ratio in a sample of $L=80$ cm is plotted in Fig.~6. It is unity in the diffusive limit, since $P_{\widehat{I}}(\widehat{\phi^{\prime}})$ is then a gaussian centered at $\langle\widehat{\phi^{\prime}}\rangle_{\widehat{I}}=1$, and rises above unity, as localization is approached. The variation with frequency of $\langle\widehat{I}\,\widehat{\phi^{\prime}}\rangle$ follows closely that of $var(\widehat{I})$ (see Fig.~2c of \cite{PRL}). The localization threshold, at which $var(\widehat{I})=7/3$ \cite{PRL}, corresponds to the condition, $\langle\widehat{I}\,\widehat{\phi^{\prime}}\rangle=1.1$, shown as the dotted line in Fig.~6.

\begin{figure}
\includegraphics[width=\columnwidth]{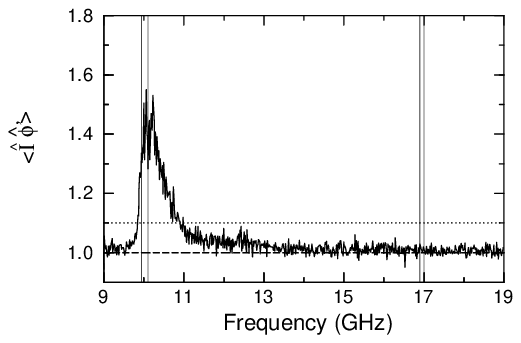}
\caption{Dimensionless ratio, $\langle\widehat{I}\,\widehat{\phi^{\prime}}\rangle\equiv\langle I_{ab}\phi^{\prime}_{ab}\rangle/\langle I_{ab}\rangle\langle\phi^{\prime}_{ab}\rangle$, versus frequency for $L=80$ cm. The dashed line corresponds to the value of unity of this ratio in the diffusive limit \cite{Bart}. The dotted line represents the condition, $\langle\widehat{I}\,\widehat{\phi^{\prime}}\rangle=1.1$, which corresponds to the localization criterion, $var(\widehat{I})=7/3$ \cite{PRL}. The peak above this line indicates the window of localization. The frequency intervals used in computing the statistics for localized and extended waves in Figs.~1-5 are marked by vertical lines.}
\end{figure}

In conclusion, we find striking differences between the statistics of dynamics of localized and extended waves. Characteristic features of the statistics of localized waves are an increasing average delay time with intensity, an asymptotic decay of the variance of the delay time proportional to $1/(\widehat{I})^{1/4}$, and a double-sided stretched exponential distribution of the weighted delay time. These features reflect transport via resonant coupling to isolated localized states. We expect that the distinctive and complex behavior observed is characteristic of electron transport as well as of propagation of all varieties of classical waves. 

\begin{acknowledgments}
We thank B.A.~van~Tiggelen for valuable discussions and gratefully acknowledge support from NSF (DMR9973959) and ARO (DAAD190010362).
\end{acknowledgments}

\end{document}